# ENRICO FERMI AND THE DISCOVERY OF NEUTRON-INDUCED RADIOACTIVITY: A PROJECT BEING CROWNED

Alberto De Gregorio[*]

ABSTRACT: This paper deals with the Physics Institute in Rome, and its getting ready for investigating neutron physics, until Fermi's discovery of neutron-induced radioactivity in 1934. The relevance of nuclear topics had been acknowledged in Rome since 1929. The Institute had been directed towards nuclear researches since then, but, still in 1933, it was not yet engaged in experimental research on nuclear physics, on account of the lack of adequate supplies. Really, an adjustment of the equipment and supplies had been undertaken, so that strong radioactive sources, Geiger-Müller counters and Wilson chambers, devoted to nuclear researches, became eventually available at the end of 1933, thanks largely to Rasetti's efforts.

On March 25, 1934 Enrico Fermi announced he had discovered neutron-induced radioactivity. Such result just represented the synthesis of two previous discoveries: that of the neutron, and that of artificial radioactivity (induced by alpha-particles, and accelerated deutons and protons). In the following October Fermi proved neutrons to be much more effective in inducing radioactivity when they were slowed down. That represented the first step towards the utilisation of nuclear energy.

Italian physics had been lagging behind for decades in comparison with the leading European countries, and the United States. Between 1933 and 1934, Fermi published the β-decay theory. Now, through his experiments of neutron-induced radioactivity, he showed the main lines of research in neutron physics. The Physics Institute in Rome eventually became a reference pole in nuclear researches.

Such a successful achievement represented the crowning of a project of renewal, devised by the Director of the Roman institute of physics, Orso M. Corbino, since mid 1920s. We are going to review here the main steps which, on the experimental side, opened the path to researches on the nucleus in Rome. Through such review the outstanding contribution clearly stands out of Franco Rasetti, who was Fermi's close collaborator and friend.

## 1. – TOWARDS THE NUCLEUS

Italian physicists began to be interested in the atomic nucleus as early as the end of the 1920s. Until then, the study of atoms and molecules had prevailed. In September 1929 Corbino, giving his address before the *Società italiana per il progresso delle scienze*, said:

> One can come to the conclusion that tremendous advancement in the ordinary domain of experimental physics is improbable, while many possibilities open in attacking the atomic nucleus, which is the real field of tomorrow's physics.[1]

---

As much aware of the future to be expected of microphysics was Fermi:

> Nowadays, that the investigation of the atom and molecule is on its way to having a solution, physicists are beginning to turn their attention, with increasing insistence, towards a new problem: the nuclear structure.[2]

Really, researches carried out by Fermi and his collaborators in Rome had concerned completely different topics until then, focusing on quantum electrodynamics on the theoretical side, and on atomic and molecular spectroscopy, on the experimental side. Only very few exceptions concerned theory. I fact, as early as 1928 Giovanni Gentile jr had dealt with a nuclear model suggested by Ernest Rutherford in 1927. Evidence of Ettore Majorana's and Fermi's interest in the nucleus dates to 1929. In July, Majorana disputed his thesis, improving some calculations by Gamow and Houtermans, and tackling the problem of radioactive nuclei which recombined by alpha-particle capture. As for Fermi, he published a paper on the theory of nuclear magnetic moments at the end of 1929.[3]

Rasetti attained important results on the experimental side during the academic year 1928-29, when he was at the California Institute of Technology in Pasadena. His success nourished interest in the nucleus in Rome. A typescript by Rasetti is kept at the Physics department of «La Sapienza» in Rome.[4] It helps in making his engagement on experimental nuclear physics come out. Rasetti specifies that «the chief purpose of these notes is to facilitate the task of those friends and colleagues who will write my obituaries [...], and to avoid those errors that almost inevitably creep in, even when writing about persons with whom one has been closely associated». Rasetti recalls that, while in Pasadena, he began investigating the Raman effect in gases, as soon as the Raman effect was discovered: «I undertook this work entirely of my own initiative, since nobody at that laboratory had any previous experience with Raman techniques». The nitrogen nucleus $^{14}_{7}N$ had already been discovered to have spin 1; Rasetti furthermore showed it to obey Bose-Einstein statistics. Both these two properties strongly challenged the model of a nitrogen nucleus consisting of protons and electrons (in odd number altogether).

Commenting on one of Fermi's «Collected papers», in 1962 Rasetti furthermore recalled:

> [Fermi's] intention of entering the nuclear field first became manifest in the winter of 1930-1931, although his theoretical work at the time followed entirely different lines. He set himself as a first task the construction and operation of a cloud chamber, with the help of E. Amaldi. The weakest point of the Rome physics department was its poorly equipped and inefficiently staffed machine shop [...]. These circumstances induced Fermi to make use of the «do-it-yourself» methods that were characteristic of him both in theoretical and experimental work. [...] Alpha-particle tracks were soon observed, but their quality was poor owing to the slowness of the expansion and difficulty of clearing the field of old tracks. [...] In the spring of 1931 Fermi gave up the cloud chamber project and went back to theoretical work.[5]

---

[1] Orso M. Corbino, «I compiti nuovi della fisica sperimentale», *Atti della Società italiana per il progresso delle scienze* (1929), p. 12.

[2] Enrico Fermi, «La fisica moderna», *Nuova Antologia 65* (1930), 137-145; also reported on: Enrico Fermi, *Note e memorie – Collected Papers* (hereafter cited as *FCP*), Edoardo Amaldi *et al.* eds., 2 vols., vol. 1: *1921-1938* (Rome – Chicago: Accademia Nazionale dei Lincei – the University of Chicago Press, 1962), pp. 371-378.

[3] Giovanni Gentile, «Sulla teoria dei satelliti di Rutherford», *Rendiconti dell'Accademia dei Lincei 7* (1928), 346-349; Ettore Majorana, «Sulla meccanica dei nuclei radioattivi», thesis (unpublished); Enrico Fermi, «Magnetic Moments of Atomic Nuclei», *Nature 125* (1930), 16.

[4] Franco Rasetti, *Biographical Notes and Scientific Work*, autobiographical note, kept at «Edoardo Amaldi» Archive, at the Physics Department of the University «La Sapienza» in Rome, 1968.

Really, those described by Rasetti should be considered just the first, uncertain steps ever made in Rome towards the experimentation upon the nucleus. In fact, the only definite results concerning nuclei which had been published by the physicists of the Institute were circumscribed to those already mentioned: those attained by Rasetti (though in Pasadena), and the theoretical ones by Gentile and Fermi. Still in 1931, the Institute was mainly engaged in 'classical' researches in atomic and molecular physics, in spite of Corbino and Fermi having realised what remarkable prospects were to be expected in nuclear physics. An international congress on nuclear physics was eventually held in Rome in October 1931, and that officially marked the time when the Roman Institute turned its activities towards nuclear topics.

## 2. – A PROJECT WELL UNDER WAY

The congress of the 1931 boosted the realisation of the nuclear programs. In particular, Rasetti spent another academic year abroad: between 1931 and 1932 he visited Lise Meitner's laboratories in Berlin-Dahlem, and there he practised the techniques proper to nuclear physics. Rasetti, taking advantage of his skilfulness, would play a fundamental role as an experimentalist in the activity and achievements by the Roman Institute. Amaldi, in an interview with Charles Weiner, clearly testified that «Segrè and I both learned to do experimental work from Rasetti. Rasetti was a very good experimentalist, and he taught us how to work. He was much better than Fermi as an experimentalist. [...] Rasetti went to Berlin, and he was working with Meitner. It was part of this idea to learn nuclear physics techniques».[6] In his autobiographical note, Rasetti recalls the reasons why he went to Germany as follows:

> Discussions with Fermi and the other physicists in Rome had led to the conclusion that spectroscopy was rapidly becoming a field where new fundamental discoveries could no longer be expected, whereas the nucleus appeared as the likely centre of interest in the near future. Hence the decision to learn and import nuclear techniques. In Dahlem I familiarised myself with the preparation of radioactive sources and the use of counters, ionisation chambers and cloud chambers.

The current standard of the laboratories in Rome had led Fermi to soon give up his attempt to handling cloud chambers. It was therefore needed to «import nuclear techniques», in order to have the capacities of the laboratories improved, and make them as efficient as many laboratories abroad were.

A few months after the congress on nuclear physics, the discovery of the neutron further spurred the Roman group to pursue the nuclear program. Rasetti learned about the neutron while in Berlin. He readily undertook experiments, showing that the penetrating radiation from beryllium, besides including neutral particles similar in mass to protons, also comprised an electromagnetic component. Rasetti resorted to the coincidences technique. He used two aluminium Geiger-Müller counters, with walls of 0.5 mm. Radiation from the polonium-beryllium source was filtered through 2 cm of lead. The number of the coincidences increased, from about 12 per hour due to the background, to 90. Rasetti then placed some aluminium shields a few millimetres thick between the counters, and the number of coincidences decreased, though neutrons, as Rasetti remarked, were not significantly absorbed through such aluminium shields. On the other hand, the coincidences could not be indirectly produced by neutrons which had projected nuclei or electrons, since the penetrating power of the latter would be too small. Rasetti therefore concluded that coincidences were produced by fast electrons, projected by Compton effect by photons of about 10 MeV. Beryllium radiation, far from being homogeneous, consisted of a «mixture» of neutrons and radiation quanta. Rasetti underlined that gamma-rays emission at beryllium disintegration was no surprise.

---

[5] *FCP* (ref. 2), 548.

[6] Edoardo Amaldi, oral history interview by Charles Weiner, April 9-10, 1969, Niels Bohr Library, American Institute of Physics, College Park, MD USA.

While in Berlin, Rasetti also investigated the intensity of beryllium radiation, as a function of the incident alpha-particles. The next year Gilberto Bernardini, he too at Lise Meitner's laboratories for a stay, confirmed Rasetti's results.

Rasetti, mastering the nuclear techniques as early as the beginning of 1932, could well bear comparison with the leading physicists in the van of experiments in nuclear physics. Rasetti's skilful handling concerned Geiger-Müller counters as well as Wilson chambers: two figures in his paper show the tracks of recoiling protons produced by neutrons in a cloud chamber.

On September 10, 1932 Fermi wrote to Segrè about the projects for the Institute: «I have no program for next year's work: I do not even know whether I shall start fooling around with the Wilson Cloud Chamber again, or if I shall again become a theoretician. […] The problem of equipping the Institute for nuclear work is certainly becoming even more urgent if we do not want to fall into a state of intellectual slumber».[7] Really, some opposition arose – particularly from Segrè – to leave a well-known field, and venture into the little known nuclear world. In fact, even if there was consciousness that the spectroscopic field was being gradually worked out, still, spectroscopy warranted a tested research field.

In any case, the lines to be followed were being fixed on. In the already recalled comment to Fermi's papers, Rasetti wrote:

> In the fall of 1932, Fermi and Rasetti organised a joint program of research in nuclear physics in Rome. To minimise the drawback of the inadequate machine shop, several instruments were designed and their construction was «farmed out» in a private shop in Rome. A rather large cloud chamber, essentially designed after those in use in Berlin-Dahlem, worked excellently as soon as it was assembled. […] These developments were made possible by a grant from the Consiglio Nazionale delle Ricerche, which had raised the research budget of the department to an amount of the order of $ 2000 to $ 3000 per year; a fabulous wealth when one considers that the average for physics departments in Italian universities was about one-tenth of that amount.

The project was gradually advancing, and the supplies being adapted. Nevertheless, only theoretical results were being achieved in Rome for some time. In fact, Fermi and Segrè dealt with the theory of hyperfine structure, Renato Einaudi with calculations on the forbidden lines related to the nuclear spin, Gian Carlo Wick reflected on some aspects of neutron-proton interaction. Moreover, in 1933 Majorana published an important theory of nuclei, entrusting a fundamental role to neutrons; at the end of the year, Fermi published the well-known beta-decay theory. Notwithstanding all those pursuits, no experimental work about neutrons was yet carried out in Rome until 1934: the only experiments on neutrons carried out by Italian physicists were those made by Rasetti and Bernardini while they were in Germany, in 1932 and 1933. In France, in England, in Germany, and in the United States, neutron investigations were in full swing, instead. Abroad, neutron scattering and absorption were investigated, nuclear transmutations induced by neutrons had been discovered, secondary electrons related to the passage of the penetrating radiation from beryllium were to be reckoned with.[8]

---

[7] Emilio Segrè, *A Mind Always in Motion. The Autobiography of Emilio Segrè* (Berkeley and Los Angeles: University of California Press, 1993), p. 86; in the letter Fermi wrote «non so nemmeno se tornerò a camera-di-wilsoneggiare», which sounds as: «neither I know whether I'm Wilson-chambering again».

In so far as we merely consider the results actually achieved, the condition of the nuclear physics practised in Italian laboratories until 1933 might resemble that proper of the Italian physics during the two first decades of the century, when Italy was impervious, if not refractory, to the wave of renewal produced by quantum and relativistic theories. Still, that is only an outward similarity. In the substance, the deep difference between the sterility at the beginning of the century, and the idleness in the early Thirties can be summarised with one only word: project. In fact, even if experimental results were still lacking, Corbino and Fermi had realised the importance of the new line of researches, and had promoted a definite though gradual adaptation of the work to that carried out abroad. The nuclear program was being accomplished. Papers from abroad were being systematically studied in Rome, and the nuclear techniques were being learned. At the end of 1933 Fermi published the β-decay theory. When the project of renewal was completed, and the favourable moment came, many fundamental results in the experimental field came as well.

## 3. –RADIOACTIVE SOURCES AND DETECTION TECHNIQUES

Radioactive sources were needed in experimental nuclear physics. In a paper published at the beginning of 1934 Rasetti clearly stated:

> An organic program, aimed to equip the Institute itself for researches in nuclear physics, is being developed at the Physics Institute in the University of Rome, with the help of the National Council of Researches. Such kind of work cannot be improvised. On the contrary, they demand complex organisation, on account of the absolutely special instruments which are proper to radioactivity techniques, and – even more – because of the difficulties in providing radioactive substances, in separating them, and in letting them be in shape in the experiments.[9]

In 1933, there were only a few milligrams of radium salts as a whole in the physics laboratories of the Italian universities. That was all the radioactive substances they owned. Even if such amount might fit for didactical demonstrations on ionising effects, it certainly did not fit for nuclear researches. Notwithstanding this, the same rooms, in which the Physics Institute of the University of Rome was, also accommodated the laboratories of the Institute of Public Health, directed by Giulio Cesare Trabacchi. What's more, the laboratories of Public Health, differently from those of the Physics Institute, owned a huge amount of uranium. According to Oscar D'Agostino,[10] the chemist of Fermi's group, there were 1,600 mg of radium salts: 1,300 mg were used for medical therapies, the radon being extracted from them and distributed to the Roman hospitals weekly. Instead, 300 mg of radium salts had remained unused «for particular events» for many years, and the Public Health had just gained them. As we are going to see soon, sources to be used in nuclear researches were eventually obtained from those 300 mg of radium salts.

---

[8] See: Alberto De Gregorio, «Neutron physics in the early 1930s», *Historical Studies in the Physical and Biological Sciences 35, n. 2* (2005).

[9] Franco Rasetti, «Sopra un forte preparato di Radio D ottenuto nell'Istituto Fisico di Roma», *La Ricerca Scientifica 5*, n. 1 (1934), 3.

[10] Oscar D'Agostino, «L'era atomica incominciò a Roma nel 1934», 1st instalment, *Candido 16*, n. 23, June 8, 1958, p. 23.

We have evidence of the use of radioactive sources for research purposes in Rome from Nella Mortara, a collaborator of Trabacchi at the Institute of radium. In a paper of 1932 she described one of the two plants (fig. 1), arranged to extract radium emanation (Rn) from radium chloride solutions containing 200 and 1041 mg of radium respectively.[11] Mortara would carry out researches on radon later on. In 1933 she improved some methods for the 'calibration' of radioactive preparations, resorting to ionisation chambers. Moreover, she investigated the diffusion properties of the radon gas, and the optimisation of methods for its purification.

Furthermore, radon was used by Fermi and Rasetti themselves, in 1933: they resorted to 100-150 mCi of radon as a gamma-source, and to a bismuth crystal as a gamma-ray spectrometer. Really, they only obtained poor results, since very strong sources would have been needed by such tool, in order to investigate gamma-rays spectra.

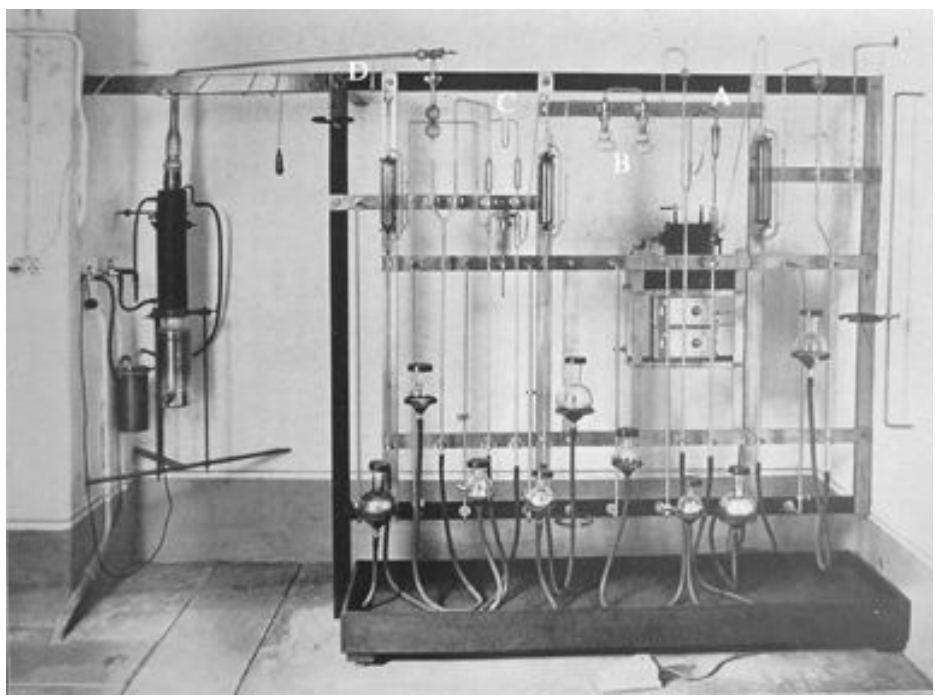

FIGURE 1. – A partial view of one of the two plants meant for radon extraction. The radium chloride solution was in a contiguous room. In the *A* tube some sparks were produced, and hydrogen and oxygen combined. Drying substances were contained in *B*. Radon liquefaction was obtained by dipping the *C* loop into liquid air. The vertical *D* tube communicated with a capillary, inside which emanation was pumped. The large bulbs at the bottom were filled with mercury, so as to open and close the various parts of the extraction circuit (from: Mortara (ref. 11)).

Instead of radon sources, polonium sources were much more preferred in nuclear researches, since they almost exclusively emitted alpha-particles. As for polonium, Rasetti eventually succeeded in the 'exploit' of gaining the new radium preparation of

---

[11] Nella Mortara, «L'Ufficio del Radio», *Rivista di radiologia e fisica medica 4*, n. 4 (1931-32), 462-468.

300 mg from Trabacchi, so as to produce polonium sources. In fact, polonium too was extracted from radium, even though more complex procedures were required than for radon. In 1934 Rasetti recalled that the 300 mg of radium salts, that the Direction of the Public Health had recently obtained, had been unused «for complicated events» for 14 years. The decay chain starting from radon went through radium D ($Pb^{210}$), radium E ($Bi^{210}$), and polonium, *i.e.* radium F ($Po^{210}$). Now, all the products of radon preceding Ra D had half-lives of the order of some minutes, while radium D had a half-life of 22 years, which meant that, in order to extract polonium, strong radium supplies were needed, from which emanation should not be extracted for many years. That was just the shape in which the 300 mg of radium salts supplied by Trabacchi were.

Rasetti had been once again at the Keiser Wilhelm Institut in Berlin-Dahlem in 1933, learning the separation techniques. The procedures he applied to a small amount of radium there, he would use again in Rome. Back to Rome, in November Rasetti separated radium D from the radium put at his disposal by Trabacchi, so that polonium could be periodically extracted from Ra D. D'Agostino helped Rasetti in the separation work. Furthermore Rasetti, being completely thrown into nuclear researches, with Corbino and Fermi's agreement let D'Agostino get a grant from the National Council of Researches, and go to Marie Curie's laboratories in Paris, in order to «practise in the manipulations proper of radioactive chemistry». Thanks to an «almost quantitative separation», a radium D preparation of about 110 mCi was finally obtained in Rome. That was «one of the strongest all over the world, perhaps second only to that at M.me Curie's laboratory».[12] So strong its activity was, that Rasetti could also send some polonium to other laboratories (in particular to Gilberto Bernardini's), where polonium might be used in neutron researches.

As we have already seen, at the end of 1929 the relevance of nuclear researches was acknowledged by the Roman Institute, while its laboratories boosted towards nuclear topics starting from about the end of 1931. Now, a clear-cut moment can be pointed out, when the experimental activity in Rome definitely turned to actual nuclear researches: it was at the end of 1933, when, having a large amount of radium salts been providentially obtained, and having Trabacchi allowed Rasetti to use it, Rasetti prepared strong polonium supplies from it, to be used in nuclear researches.

Until 1933, the laboratories in Rome had been only engaged in spectroscopy researches, on the one side, and in the adjustment to nuclear purposes, on the other side. Still, some researches on weak radioactive substances had been carried out in Arcetri, near Florence, even if much poorer economical resources were available there. Such researches were not in the van, nor concerned nuclear physics properly speaking. Still, the adopted techniques, and moreover some of the substances investigated, made them similar, in some relevant respects, to the researches which were soon to be carried out on neutron-induced radioactivity in Rome.

Daria Bocciarelli, Gilberto Bernardini, Giuseppe Occhialini, Giulio Racah, Bruno Rossi were some of the young researchers of the Physics Institute of the Univesity of Florence situated in Arcetri, at the end of the 1920s. Rossi undertook the manufacturing of some geigers and of the related electronics between the autumn of 1929 and the following spring. In the summer of 1930 he went to Berlin for some months, at Bothe's laboratory, and there he performed a modified version of Bothe and Köhlorster's experiment on the nature of cosmic rays, by resorting to Geiger-Müller counters. Rossi's deep interest in geomagnetic effects also arose while he was there, and Rossi soon became one of the leading researchers in cosmic rays physics.

---

[12] Rasetti (ref. 9), p. 5. The 110 mCi of radium D separated by Rasetti should actually correspond to an almost complete separation. In fact, radium D has a half-life of 22 years; therefore, $300 \cdot [1 - \exp(-14/22 \cdot ln2)]$ mCi = 107 mCi of radium D are produced from 300 mg of radium in 14 years. Still, Mortara referred that the chloride solution contained only 200 mg of radium, not 300 mg (see: MORTARA, (ref. 11)).

When Occhialini and Bocciarelli closely investigated weak radioactive substances in Arcetri, Rossi provided them with precious advice for the devices to be used, comprising Geiger-Müller counters. A systematic difficulty arose in such experiments: the very weak activity of the preparations prevented researchers from using ordinary methods in the magnetic analysis of the emitted beta-rays. Ordinary methods implied the resort to small sources. Elements with such a weak activity, as rubidium and potassium have, should necessary have a large surface instead, otherwise their activity would have not been measurable. The solution suggested by Rossi to Occhialini lay in a spectrograph, made up of a small Geiger-Müller counter with aluminium walls 7 $\mu$m thick. A cylindrical surface lined with rubidium chloride was placed co-axially round the geiger. Its radius exceeded ten times that of the geiger (fig. 2). A telephonic counter registered the signals, produced by the Geiger-Müller counter and then amplified. A magnetic field was finally applied parallel to the cylinders axes. As the intensity of the field increased, fewer and fewer electrons could reach the geiger, thus allowing one to determine the electronic spectrum.

While investigating potassium beta-spectrum, Daria Bocciarelli resorted to the same device used by Occhialini for rubidium, and to a solution of potassium chloride. Note that Fermi himself presented one of Bocciarelli's communications to the *Accademia dei Lincei*. Bocciarelli closely investigated potassium radioactivity with the coincidences method too, resorting to various interposed absorbers, and to two counters with 7 $\mu$m aluminium walls, kept at a low pressure inside a glass shield. She also used one single Geiger-Müller counter as a tool: in that case, she put for example a cylindrical surface, lined with potassium salts, round a counter with 2/10 mm aluminium walls, so that some interposed cylindrical shields could eventually lead to the absorption spectrum.

An experimental setting similar, to some respect, to Occhialini's and Bocciarelli's had already been used by Georg von Hevesy, W. Seith, and M. Pahl. In the autumn of 1931, they sent to the *Zeitschrift für physikalische Chemie* a paper reporting on their measurements of the weak radioactivity of potassium. The investigated substance was ground into powder, and then pressed as a hollow cylinder, so as to be slid on a runner, and placed co-axially around the counter (fig. 2).

The problem of investigating weak radioactivity was exactly the same that Fermi would face in his experiments on neutron induced radioactivity. Also the devices he used would be exactly those just described: Fermi's devices consisted of small Geiger-Müller counters with 'thin' aluminium walls; when possible, the specimens he investigated were hollow cylinders, so as to maximise the efficiency in activating them, and in counting the decay electrons.

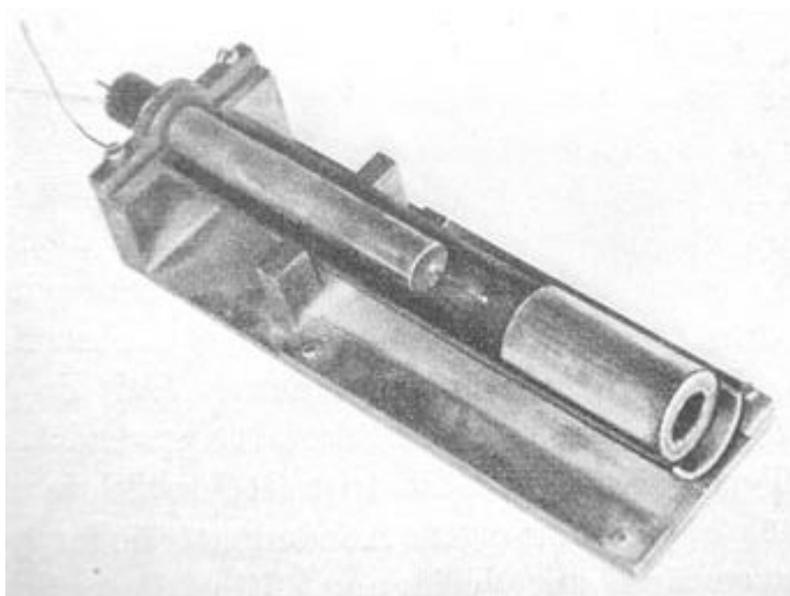

FIGURE 2. – The detection equipment used by von Hevesy, Seith, and Pahl in investigating potassium radioactivity. The Geiger-Müller counter, and the specimen as a hollow cylinder, are clearly recognisable (from: G. von Hevesy, W. Seith, M. Pahl, «Über die Radioaktivität des Kalium», *Zeitschrift für physikalische Chemie*, Bodenstein-Festband (1931), 309-318).

## 4. – HOW FERMI AND HIS GROUP REACTED TO THE DISCOVERY OF ARTIFICIAL RADIOACTIVITY

During 1933, the Physics Institute of Rome kept itself updated on nuclear researches carried out abroad. Fermi was acquainted with results achieved in other laboratories. He took part, for example, in both the meeting of the American physical society, and in the seventh Solvay conference, held in June and in October respectively.[13] Among the physicists attending in the Chicago meeting, Cockcroft, Harkins, Henderson, Lawrence, Lewis, and Livingston reviewed both researches in progress, and the results already attained in their laboratories – being about to tackle radioactivity induced by accelerated ions within some months –. Some of the topics discussed in Brussels regarded disintegration processes, produced by α-particles as well as by accelerated particles. Many questions of fundamental physics, ranging over the constitution of the neutron as well as the constitution of nuclei, were also reviewed. It was at this time that, for example, Francis Perrin foresaw the artificial radioactivity produced by α-particles. Fermi got prodded by his participation in the Solvay meeting, and he published the beta-decay theory some weeks later.

When the discovery of artificial radioactivity was reported in Rome, at the beginning of 1934, the Physics Institute was eventually ready to start a new course of researches. The foundation «Oscar D'Agostino» in Avellino, Italy, keeps a really

---

[13] *Proceedings of the American Phisical Society*, (19-24 giugno 1933), reported on Physical Review *44* (1933), 313-331; «Structure et propriétés des noyaux atomiques», proceedings of the Seventh Solvay Conference, Brussels, October 22-29, 1933 (Paris: Gauthier-Villars, 1934).

remarkable letter, that Fermi and Rasetti wrote to D'Agostino on February 9, 1934.[14] They reported on progress on nuclear techniques. The polonium supply, obtained by Rasetti with D'Agostino's help, was very clean, «giving a good fan of alpha particles, all of the same length». Fermi and Rasetti furthermore gave news that a Wilson chamber was ready, and the construction of another one was to be completed soon. The letter is here translated:

Rome, February 9, 1934

Dear D'Agostino,

we thank you for the news from you, and we are glad to hear that you could easily get your bearings in M.me Curie's Institute.

The procedure for preparing the bismuth plate is described in detail, in the note of which we are sending you two excerpts; in any case we are available for whatever further explanation. We are also sending you the photograph of a $\gamma$-rays spectrum, since reproduction is not very clear on Ricerca Scientifica. We are posting you, for the Joliots as well, a bismuth monocrystal, with one side already prepared for spectrography.

The polonium supply put in the Wilson chamber is very clean, giving a good fan of alpha particles, all of the same length; the large chamber works properly, but for the puttied parts. The other wire gauze-chamber of the Wilson type is at an advanced stage of construction. We are preparing some counters, in order to repeat Joliot's experiments on artificial radioactivity with positron emission, and to inquire whether it is possible to separate the unstable radioactive product which should be produced, within the few minutes of its mean-life.

Many wishes and greetings; please return our greetings to all our mutual acquaintances

Enrico Fermi
F. Rasetti

Thanks for the postcards and many warmest greetings
E. Segrè
E. Amaldi

It should be stressed that Fermi and Rasetti wrote that they were about «to repeat Joliot's experiments on artificial radioactivity with positron emission». It is very remarkable in fact that, according to the letter to D'Agostino, Fermi and Rasetti did not promptly react by undertaking novel researches on neutron-induced radioactivity. Rather, they only planned to repeat Joliot's experiments on alpha-induced radioactivity,[15] when news of artificial radioactivity was received in Rome!

---

[14] The Technical institute «D'Agostino» in Avellino, Italy, keeps several documents belonged to the chemist of Fermi's group, Oscar D'Agostino, who was native of Avellino. Noticeably, it was realised in 2002 that a notebook, along with a few spare sheets, largely consisted of laboratory notes by Fermi himself on the early experiments on neutron-induced radioactivity. As for the recovery of Fermi's notes, see: Giovanni Acocella, Francesco Guerra, and Nadia Robotti, "Enrico Fermi's discovery of neutron-induced artificial radioactivity: the recovery of his first laboratory notebook", *Physics in perspective*, 6 (2004), 29-41. As for Fermi and Rasetti's letter, it is reproduced in: Giulio Pugliese, *Oscar D'Agostino. Il chimico del gruppo di via Panisperna* (Avellino: Pergola Editore, 1988), pp. 77-78.

[15] One should bear in mind, however, that we cannot exclude the possibility that, on February 9, Fermi and Rasetti already had the intention of searching for neutron-induced radioactivity but that, fearing competition with the Joliot-Curies, simply preferred not to say a word of it to D'Agostino. Still, such possibility seems little likely (see De Gregorio (ref. 8)).

Fermi and Rasetti's letter shows the final steps of the Institute towards experimental nuclear physics. As we have already mentioned, in 1962 Rasetti recalled that a cloud chamber, «essentially designed after those in use in Berlin-Dahlem, worked excellently as soon as it was assembled». That suggests Rasetti's role in proposing the project for a Wilson chamber, as he came back from Berlin-Dahlem. Moreover, the letter to D'Agostino suggests that the Wilson chamber might become available to the Institute at the beginning of 1934, indicatively between D'Agostino's departure and the date of the letter.

The date when the Institute got the two chambers can be identified almost exactly: a bulky book-keeping register is kept at the University «La Sapienza», reporting that two Wilson chambers were manufactured by a firm, in accordance with Rasetti's version, and were loaded by the Institute on December 9, 1933 (fig. 3).

FIGURE 3. – The book-keeping register, reporting the two Wilson chambers and a motor for the use (Museum of Physics of the University «La Sapienza» of Rome).

## 5. – RASETTI'S ROLE

Fermi's enterprise, and his skill in manufacturing the counters on his own before discovering neutron-induced radioactivity, have often been pointed out. Really, such version lies on two accounts. For what concerned the manufacturing of the counters, in 1954 Laura Capon Fermi wrote (not without some ambiguity) that «in 1934 […] the only way to obtain Geiger counters was to make them, and Enrico did not know how»; on the other hand she added that, as Rasetti left for Morocco, «Enrico tackled the task of making counters by himself, and he had them ready in a reasonable time». The other account is by Segrè, who in 1955 agreed that Fermi «built with his own hands some primitive Geiger-Müller counters of aluminium».[16] Notwithstanding these two accounts, the letter to D'Agostino shows that some counters had been prepared by Fermi and Rasetti more than one month before neutron-induced radioactivity was discovered.

Rasetti's efforts in providing the Roman Institute with the geigers agree with what Amaldi stated in his cited interview with Weiner, pointing out the fundamental role that Rasetti played in the experimental activity carried out in Rome. Fermi himself, concerning his experiments on neutrons, fairly surprisingly referred: «I should acknowledge that I had been mainly a theoretical physicist, until that time, and largely even after then. […] As experiments become a bit more complex, they are beyond my experimental abilities […]. Therefore, those friends of mine could do what I was not able to».[17] These remarks, along with Rasetti's contribution in building the first geigers, and his directions in designing the Wilson chamber, all confirm the fundamental role that Rasetti indirectly played in the discovery of neutron-induced radioactivity, for he adjusted the equipment and supplies at the Institute. As Rasetti put it: «I do not believe that Fermi would have attempted his historical experiment unless neutron sources, counters, and other nuclear instruments had been already familiar to him».

---

[16] Laura Fermi, *Atoms in the family. My life with Enrico Fermi* (Chicago: The University of Chicago Press, 1954), 84-85; see also Emilio Segrè, in *FCP* (ref. 2), p. 640.

No other documentary proof, apart from the letter to D'Agostino, exists at present, of the fact that in February Fermi and Rasetti were to repeat the experiments carried out in Paris, on alpha-induced radioactivity with positrons emission. That draws our attention to the need of integrating the personal recollections of those who took part in those events occurred in Rome, with archive records. Really, Fermi's switch from alpha-induced radioactivity to neutron-induced radioactivity is examined closely in another paper.[18] What should be stressed here is that the adjustment of the supplies, devoted to researches on artificial radioactivity, was being completed in the Physics Institute in Rome on February 9, 1934. At that date, the laboratories were already provided with polonium sources, at least one Wilson chamber was operating, and Geiger-Müller counters were being prepared.

Historical sources generally agree that Fermi's search for neutron-induced radioactivity dated back to March 1934. Still, some accounts disagree about the first attempts in obtaining artificial radioactivity by neutrons. Not only did Laura Capon and Segrè state that Fermi built the geigers by his own just before undertaking his historical experiments, but they also stated that Fermi undertook his search for neutron-induced radioactivity on his own, resorting to a radon-beryllium source provided by Trabacchi. The intense γ-radiation emitted by radon sources was no drawback, since the measurements followed the irradiation. Segrè confirmed such version also in 1962, in his comments reported on Fermi's collected papers. Still, it is rather remarkable that, in that same recollection, Rasetti's account differed from Segrè's: Rasetti recalled that Fermi suggested to him that they two should undertake experiments with neutrons together; according to Rasetti, moreover, the first neutron source he and Fermi used was of the polonium-beryllium type. Thus, the same volume of Fermi's collected papers reports two conflicting versions, only a few pages apart the one from the other.

Also remarkable is that all the accounts to come conformed to Rasetti's version, as Fermi's collected papers were published. Segrè himself, during the Tenth International Congress of the History of Science, later in 1962 stated that «after the discovery of the alpha-induced radioactivity had been announced by Curie and Joliot, Fermi suggested to Rasetti that they try to observe similar effects with neutrons».[19] The first attempts were carried out with polonium-beryllium sources, according to Segrè's 'up-dated' account. In 1978, Segrè even stated that Fermi achieved successful results by resorting to «sources that Rasetti had prepared with the radium of the Public Health»:[20] clearly, such sources should contain the polonium, that Rasetti had extracted with D'Agostino's help at the end of 1933. In 1984, Amaldi as well agreed with Rasetti's account.

---

[17] Enrico Fermi, *Conferenze di fisica atomica*, seventh conference: «Il neutrone» (Rome: Accademia dei Lincei, 1950), p. 92; also reported in *FCP* (ref. 2), vol. 2: *1939-1954*, p. 758.

[18] De Gregorio (ref. 8).

[19] Emilio Segrè, «The Consequences of the Discovery of the Neutron», in *Proceedings of the Tenth International Congress of the History of Science* (Paris: Hermann, 1964), pp. 149-154.

Some more confirmations exist, that polonium was first used to irradiate beryllium so to obtain neutron sources. The letter to D'Agostino, and Rasetti's paper on the separation of radium D, are certainly free from distortions which might have been introduced with time. It is reported there that polonium supplies were in part used to test the 'large' cloud chamber, and in part had been given to Bernardini «for researches on neutrons». Furthermore Zaira Ollano, in Cagliari, having investigated the absorption of beryllium radiation, thanked Rasetti in October for having provided her with a polonium supply.[21] These episodes all suggest that it is absolutely likely that the polonium prepared by Rasetti was used at first, as neutron sources were manufactured in Rome.

Likewise, it is absolutely likely that Fermi turned to Rasetti, as he resolved to go through artificial radioactivity. In fact, Rasetti was unanimously acknowledged as the most talented experimental physicist in Rome. It had been Rasetti, who went twice to Berlin to learn nuclear techniques, within the scope of adjusting the Institute for nuclear researches. Moreover, Rasetti had dealt with experimental neutron physics since as early as the beginning of 1932. He took part in the manufacturing of the first Geiger-Müller counters the Roman laboratory was supplied with. He also prepared the polonium supplies, and he clearly gave directions to build a Wilson chamber «essentially designed after those in use in Berlin-Dahlem». If one considers all that, it comes out absolutely reasonable that Fermi, being «mainly a theoretical physicist», involved such a skilful colleague – who had been his friend since they were students in Pisa – when he resolved to turn into an experimental physicist (still, their first attempts with a Po + Be source apparently were unfruitful).

Fermi announced the discovery of neutron-induced radioactivity in a short note, on March 25, 1934, while Rasetti was in Morocco. Fermi had eventually succeeded in obtaining artificial radioactivity, resorting to a radon-beryllium neutron source. The latter consisted of a glass bulb, filled with beryllium powder and with 50 mCi of radium emanation, supplied by Trabacchi. A very strong $\gamma$-radiation was emitted along with neutrons, but it was no disturb, for the very fact that the phenomena searched for were subsequent to irradiation. Fermi, as he had attained a first set up of the detection devices, tackled the question of the sensitivity of the Geiger-Müller counters, choosing a substance with a very weak $\beta$-activity chosen as a standard, that is a solution of potassium-chloride (just the same kind of substance investigated by Bocciarelli in Arcetri). Fermi activated aluminium and fluorine. An aluminium cylinder was placed at first around the neutron source, to be irradiated, and then around the geiger, to be investigated. Calcium fluoride was irradiated, and then placed near the counter. Also platinum, lead, and mercurous chloride were investigated, and the reasons leading Fermi to look for neutron-induced radioactivity in them were straightforward.[22]

---

[20] Emilio Segrè, «Per il settantesimo compleanno di Edoardo Amaldi», *Il Giornale di Fisica 20* (1979), 181.

[21] Zaira Ollano, «Emissione secondaria da elementi di peso atomico medio sotto l'azione della radiazione da Po + Be», *Ric. Scient. 5*, n. 2 (1934), 374-376.

[22] For a closer analysis of Fermi's discovery of neutron-induced radioactivity, see De Gregorio (ref. 8).

CONCLUSIONS

What we have here reviewed clearly points out that the fundamental experimental discoveries made in Rome in 1934 were the result of an exacting program, strongly pursued by the director of the Physics Institute, Corbino. That project comprised the adjustment of the Institute to the requirements of experimental nuclear physics. We have seen in the present paper the ways through which the Roman Physics Institute was gradually introduced into experimental nuclear physics. As the favourable time came, the discovery of neutron-induced radioactivity could then creep in Rome. Two kinds of documents are available, concerning the first steps towards such achievement: the reminiscences of the physicists who took part in those researches, and the original documents. That allows us to compare various sources, so that gaps among them can possibly be filled, and possible inconsistencies pointed out. As it has already been suggested,[23] a historical review of the discoveries which took place in Rome should not be restricted to the recollections given by those who took part in them, reporting about their researches decades later. Really, a documentary confirmation is needed for such recollections. In particular, Rasetti's role in the path leading to Fermi's discovery of neutron-induced radioactivity clearly comes out from a close review of the various historical sources available. Not only did Rasetti strongly contribute in providing the Roman Institute with Geiger-Müller counters and Wilson chambers, but he also prepared the polonium supplies that he and Fermi almost certainly used in their first attempts of obtaining neutron-induced radioactivity.

I'm grateful to Fabio Sebastiani, of the University «La Sapienza» of Rome, for his constant advice and encouragement in the preparation of the present work. I should also thank Michele Cardellicchio, president of «D'Agostino» foundation, for his availability for consultation of the documents kept in Avellino.

---

[23] Alberto De Gregorio, «Sulla scoperta delle proprietà delle sostanze idrogenate di accrescere la radioattività indotta dai neutroni», *Il Nuovo Saggiatore 19*, ns. 3-4 (2003), 41-47; the translation into English is available at the following URL: http://arxiv.org/abs/physics/0309046.